\documentclass[12pt]{article}
\usepackage[paperwidth=216mm, paperheight=279mm, margin=1.87cm]{geometry}

\usepackage{natbib}  
\usepackage{caption}
\usepackage{setspace}
\usepackage{authblk}
\usepackage{hyperref} 
\usepackage{indentfirst}
\usepackage{mathtools}
\usepackage{amssymb}
\usepackage{amsmath}
\usepackage{booktabs} 
\usepackage{graphicx}
\usepackage{algorithm}
\usepackage{algpseudocode}
\usepackage{tabularx}
\usepackage{booktabs}
\newcolumntype{Y}{>{\centering\arraybackslash}X}
\usepackage{bbm}

\providecommand{\keywords}[1]
{
  \small	
  \textbf{\textit{Keywords:}} #1
}

\makeatletter
\def\@seccntformat#1{\@ifundefined{#1@cntformat}%
   {\csname the#1\endcsname\quad}
   {\csname #1@cntformat\endcsname}
}
\makeatother

\usepackage{amsthm}

\usepackage{theoremref}


\usepackage[normalem]{ulem}

\title{Sample size and power determination for assessing overall SNP effects in joint modeling of longitudinal and time-to-event data}

\author[1,2,3]{Yuan Bian}
\author[2,3]{Shelley B. Bull}
\affil[1]{Department of Biostatistics, Columbia University, United States}
\affil[2]{Division of Biostatistics, Dalla Lana School of Public Health, University of Toronto, Canada}
\affil[3]{Lunenfeld-Tanenbaum Research Institute, Sinai Health, Canada}
\date{}

\begin{document}
\maketitle

\begin{abstract}
Longitudinal biomarkers are frequently collected in clinical studies due to their strong association with time-to-event outcomes. While considerable progress has been made in methods for jointly modeling longitudinal and survival data, comparatively little attention has been paid to statistical design considerations, particularly sample size and power calculations, in genetic studies. Yet, appropriate sample size estimation is essential for ensuring adequate power and valid inference. Genetic variants may influence event risk through both direct effects and indirect effects mediated by longitudinal biomarkers. In this paper, we derive a closed-form sample size formula for testing the overall effect of a single nucleotide polymorphism within a joint modeling framework. Simulation studies demonstrate that the proposed formula yields accurate and robust performance in finite samples. We illustrate the practical utility of our method using data from the Diabetes Control and Complications Trial.
\end{abstract}


\section{Introduction}\label{intro}
Diabetes, characterized by chronically elevated blood glucose levels, is among the most prevalent and rapidly increasing chronic diseases worldwide. In $2021$, an estimated $38.4$ million Americans, approximately $11.6$\% of the U.S. population, were living with diabetes \citep{cdc2021}. Globally, the number of adults with diabetes is projected to reach $693$ million by $2045$, a more than $50$\% increase from 2017 levels \citep{cho2018idf}. As a leading cause of death, diabetes is associated with a wide range of long-term complications, including microvascular conditions (e.g., nephropathy, retinopathy, and neuropathy) and macrovascular diseases (e.g., cardiovascular disease and stroke) \citep{papatheodorou2018complications}. These complications contribute substantially to increased mortality, blindness, kidney failure, and reduced quality of life \citep{morrish2001mortality}. Type $1$ diabetes, also known as insulin-dependent diabetes mellitus, results from the pancreas's inability to produce sufficient insulin. Individuals with type $1$ diabetes require lifelong insulin therapy to maintain normoglycemia \citep{chiang2014type}. Although its precise etiology remains unclear, it is widely believed to involve complex interactions between genetic and environmental factors \citep{cole2020genetics}.

Because poor glycemic control is a major risk factor for diabetes-related complications \citep{knuiman1986prevalence}, a key scientific question is whether genetic variants influence the timing of these complications through glycemic pathways, typically measured via hemoglobin A1c (HbA1c), or through alternative mechanisms. Addressing this question requires a study design with adequate statistical power, as sample size directly impacts the precision of parameter estimates and the ability to detect meaningful genetic effects. However, practical constraints often limit data collection, making rigorous sample size estimation a critical component of study design.

Several statistical approaches have been developed for sample size estimation in survival analysis. \citet{schoenfeld1983} introduced a widely used formula based on the Cox proportional hazards model for comparing two randomized groups. Other contributions include log-rank test–based approaches \citep{freedman1982tables, lakatos1988sample}, and methods for non-binary covariates under exponential survival assumptions \citep{zhen1994sample}. \citet{Hsieh2000} extended Schoenfeld’s formula to accommodate continuous and categorical covariates without assuming an exponential survival distribution. More recently, \citet{Chen2011SampleData} proposed sample size formulas for the associations between longitudinal biomarkers and survival outcomes, as well as overall treatment effect. \citet{Wang2014} developed sample size formula to incorporate time-dependent covariates in proportional hazards models.

In genetic studies, the exposure variable is often a single nucleotide polymorphism (SNP), typically coded as a dosage variable with three levels (e.g., $0$, $1$, or $2$ copies of a risk allele). Depending on the minor allele frequency, genotype distribution may be highly unbalanced, potentially reducing statistical efficiency and affecting the validity of asymptotic approximations. Moreover, unlike treatment effects, which are generally assumed to be directional (beneficial or harmful), genetic effects can increase or decrease the risk or timing of an event, necessitating two-sided hypothesis testing.

In this paper, we derive a closed-form sample size formula for testing the overall SNP effect within a joint modeling framework for longitudinal and survival data. We further developed an interactive Shiny app to determine sample size and power, available at \url{https:// krisyuanbian.shinyapps.io/PowerSNP/}. Our method extends the approach developed by \citet{Chen2011SampleData} to accommodate non-binary covariates such as SNPs. We illustrate the utility of our approach using data from the Diabetes Control and Complications Trial \citep[DCCT;][]{TheDCCTResearchGroup1986, TheDCCTResearchGroup1990}, a multicenter randomized controlled clinical trial initiated in 1983 that enrolled 1,441 individuals with type 1 diabetes. Participants were randomized to receive either intensive insulin therapy or conventional treatment and were followed for an average of 6.5 years. The study collected rich longitudinal data on glycemic control (e.g., quarterly HbA1c measurements) and detailed time-to-event data on diabetic complications such as retinopathy. A genome-wide association study (GWAS) by \citet{Paterson2010} identified several SNPs associated with these complications in both treatment arms. Our analysis suggests that for either treatment arm, the DCCT sample size is insufficient to achieve adequate power at conventional GWAS significance levels.

The remainder of this paper is organized as follows. Section~\ref{sec: nf} introduces the joint modeling framework and presents our new sample size formula for assessing overall SNP effects. Section~\ref{sec: ss} evaluates the accuracy and robustness of the proposed formula under varying parameter settings and visit schedules. In Section~\ref{sec: a}, we apply the methodology to the DCCT data and demonstrate its practical utility. We conclude with a discussion in Section~\ref{sec: d}.

\section{Methods} \label{sec: nf}
\subsection{Preliminaries and Notation}
For subject $i=1,\ldots,n$, let $\tilde{T}_i$ and $C_i$ denote the event and censoring times, respectively. Define the observed time as $T_i=\min(\tilde{T}_i,C_i)$ and the event indicator as $\Delta_i = \mathbbm{1}(\tilde{T}_i\leq C_i)$. Let $\text{SNP}_i$ represent the genotype for a selected single nucleotide polymorphism (SNP), coded as $\{0, 1, 2\}$, corresponding to homozygous reference, heterozygous, and homozygous minor genotypes.  Under the assumption of Hardy–Weinberg equilibrium, the genotype frequencies follow $(q^2, 2pq, p^2)$, where $p$ and $q=(1-p)$ denote the minor and major allele frequencies, respectively. Let $Y_i(t)$ denote the observed longitudinal measurement at time $t\geq0$, recorded intermittently up to $T_i$. This process is subject to measurement error, with $\eta_i(t)$ representing the unobserved true underlying trajectory.

The joint modeling framework links two sub-models, one for the longitudinal process $Y_i(t)$ and one for the time-to-event outcome $T_i$, by incorporating the true longitudinal trajectory into the hazard function. For the survival process, we assume a proportional hazards model of the form:
\begin{equation}
\label{eq: phm}
    \lambda_i(t) = \lambda_0(t) \exp \left\{ \gamma_g \text{SNP}_i + \alpha \eta_i(t) \right\},
\end{equation}
where \(\lambda_0(t)\) is the baseline hazard, \(\gamma_g\) represents the direct effect of the SNP on the hazard, and \( \alpha \) quantifies the association between the true longitudinal trajectory and the event risk.

The observed longitudinal outcome is modeled as:
\begin{equation*}
Y_i(t) = \eta_i(t) + \epsilon_i(t)
\end{equation*}
where $\epsilon_i(t)\sim N(0, \sigma_e^2)$ represents independent measurement errors across time and subjects. The true trajectory $\eta_i(t)$ is modeled as a subject-specific polynomial function of time:
\begin{equation}
\label{eq: eta}
\eta_i(t) = (\beta_0 + b_{i0}) + (\beta_1 + b_{i1}) t + (\beta_2 + b_{i2}) t^2 + \cdots + (\beta_q + b_{iq}) t^q + \beta_g \text{SNP}_i,
\end{equation}
where $\boldsymbol \beta_i = (\beta_0, \beta_1, \cdots, \beta_q, \beta_g)^\top$ are fixed effects and $\boldsymbol b_i = (b_{i0}, b_{i1}, \cdots, b_{iq}^\top) \sim N(0, \Sigma)$ are subject-specific random effects capturing individual deviations from the population-level trajectory. The random effects are assumed to be independent of the measurement error.

Substituting \eqref{eq: eta} into the hazard model yields:
\begin{equation*}
    \lambda_i(t) = \lambda_0(t) \exp \left[ (\gamma_g + \alpha \beta_g) \text{SNP}_i + \alpha \left\{(\beta_0 + b_{i0}) + (\beta_1 + b_{i1}) t + \cdots + (\beta_q + b_{iq}) t^q \right\} \right].
\end{equation*}
In this framework, $\beta_g$ reflects the SNP's effect on the longitudinal trajectory, while $\gamma_g$ captures its direct effect on the hazard. The term $\alpha\beta_g$ represents the SNP's indirect effect on the hazard mediated through the longitudinal process. Thus, $\gamma_g + \alpha \beta_g$ quantifies the SNP's overall effect on event risk, combining both direct and indirect (longitudinally mediated) contributions. Our primary interest lies in testing this overall effect, i.e., whether $\gamma_g + \alpha \beta_g=0$.

\subsection{Sample Size and Power Determination for the Overall SNP Effect} \label{subs: sso}
To assess the overall SNP effect, we test the null hypothesis $H_0: \gamma_g + \alpha \beta_g = 0$ against the alternative $H_a: \gamma_g + \alpha \beta_g \ne 0$. While \cite{Chen2011SampleData} proposed a sample size formula for detecting overall treatment effects in joint models, SNPs present unique challenges: they are categorical with three levels (e.g., $0$, $1$, or $2$ copies of the minor allele) and bidirectional in nature. Consequently, a two-sided test is more appropriate when evaluating genetic effects.

Following \cite{schoenfeld1983} and \cite{Chen2011SampleData}, the number of events $D$ required to achieve power $1-\tilde{\beta}$ at a two-sided significance level $\tilde{\alpha}$ is given by:
\begin{equation}
\label{eq: ss2}
D=\frac{{(Z_{1-\tilde{\beta}} + Z_{1-\tilde{\alpha}/2})}^2}{2pq(\gamma_g +\alpha\beta_g)^2},    
\end{equation}
where $Z_{1-\tilde{\alpha}/2}$ and $Z_{1-\tilde{\beta}}$ are quantiles of the standard normal distribution corresponding to the desired significance level and power, respectively.

This formula parallels the classical treatment-effect sample size expression, substituting the binary treatment variance with the SNP genotype variance $2pq$, and adjusting for two-sided testing by using $\tilde{\alpha}/2$. It provides a practical tool for determining the number of events required to detect SNP effects of varying magnitude and allele frequencies. As shown in \eqref{eq: ss2}, the required number of events $D$ is inversely proportional to both the genotype variance $2pq$ and the squared overall SNP effect size $(\gamma_g +\alpha\beta_g)^2$. Consequently, SNPs with lower minor allele frequencies (i.e., smaller $p$) or smaller overall effect sizes necessitate larger sample sizes to achieve adequate power. These relationships highlight the importance of accounting for both allele frequency and effect magnitude when assessing study feasibility, especially when targeting rare variants or modest genetic effects.

To compute the power for given numbers of events and effect size, we invert the sample size formula:
\begin{equation}
\label{eq: p}
1 - \tilde{\beta} = \Phi \left( \sqrt{2pqD}(\gamma_g + \alpha \beta_g) - Z_{1-\tilde{\alpha}/2} \right)    
\end{equation}
where $\Phi(\cdot)$ denotes the cumulative distribution function of the standard normal distribution. \eqref{eq: p} shows that power increases with larger values of $D$, $\gamma_g +\alpha\beta_g$, or genotype variance $2pq$, and decreases as the effect size or allele frequency diminishes. In addition, a more stringent significance threshold, such as when correcting for multiple comparisons in genome-wide studies, increases $Z_{1-\tilde{\alpha}/2}$, thereby reducing power for a fixed sample size. These trade-offs underscore the need to balance power and false positive control, particularly in studies designed to detect modest genetic effects or rare variants. When $D$ is unobserved in simulations, we approximate power by replacing $D$ with its empirical estimate $\bar{D}$, the average number of events observed.

\subsection{Calculating Empirical power for the Overall SNP Effect}
To validate the calculated power, we estimate the empirical power as the proportion of simulation replicates in which the null hypothesis is rejected at the significance level \( \tilde{\alpha} \) (that is, when the p-value is less than or equal to \( \tilde{\alpha} \)). Direct joint modeling (JM) approaches for estimating and testing the overall SNP effect can be computationally intensive and challenging, as they typically rely on the Expectation-Maximization (EM) algorithm \citep{Rizopoulos2012JointR}. These challenges are compounded as the number of time points or sample size increases, leading to substantially longer computation times. In light of these limitations, two-stage methods provide a practical alternative to JM by significantly reducing computational burden, particularly in genome-wide settings \citep{canouil2018jointly}. Based on these considerations, we propose the following two-stage approach to estimate the overall SNP effect.

To evaluate the overall SNP effect in practice, we first fit the full longitudinal model to estimate \( \beta_g \). Using this estimate, we compute adjusted longitudinal responses as \( y'_i(t) = y_i(t) - \hat{\beta}_g \text{SNP}_i \), effectively removing the SNP-related variation from the observed measurements. We then fit a reduced longitudinal model of the form:  \( y'_i(t) = \eta_i'(t) + \epsilon'_i(t)\) to estimate the adjusted trajectory \( \eta_i'(t) \),  which is independent of SNP effects. Next, we fit a survival model incorporating the adjusted trajectory: \( \lambda_i(t) = \lambda_0(t) \exp \left\{ \theta_g \text{SNP}_i + \alpha' \hat{\eta}'_i(t) \right\}\), where \( \theta_g \) estimates the overall SNP effect on the hazard, and \( \alpha' \) quantifies the association between the adjusted longitudinal trajectory and survival risk. The significance of the overall SNP effect is assessed under the null hypothesis \( H_0: \theta_g = 0 \), and empirical power is calculated based on the proportion of rejections across simulated datasets.

\section{Simulation Studies}\label{sec: ss}
We evaluate the finite-sample performance of the proposed sample size formula through simulation studies. A range of simulated data is used to assess its accuracy in estimating required sample sizes under varying conditions, including different power levels, significance thresholds, visit schedule, random effect, and SNP effect sizes. For each parameter configuration described below, $1000$ simulations are conducted with a fixed sample size of $1000$.

\subsection{Simulation Design and Data Generation}
We begin by discretizing time over a prespecified grid $t_{\text{grid}}$. For subject $i$, we simulate the genotype variable SNP$_i$ from the set $\{0, 1, 2\}$ with probabilities $(0.49, 0.42, 0.09)$, respectively. For the longitudinal process, subject-specific random effects $\boldsymbol b_i = (b_{i0}, b_{i1})$ are drawn from a bivariate normal distribution with mean zero and covariance matrix $\Sigma$. The covariance matrix $\Sigma$ is parameterized by $\delta^2_{0}$, $\delta^2_{1}$, and $\delta_{0, 1}$, representing the variance of the random intercept, the variance of the random slope, and their covariance, respectively. The true underlying longitudinal trajectory for subject $i$ at time $t$ is given by
\[
\eta_i(t) = (\beta_0 + b_{i0}) + (\beta_1 + b_{i1}) t + \beta_2 t^2 + \beta_g \cdot \text{SNP}_i.
\]
This function is evaluated at all scheduled time points $t_{\text{grid}}$. Observed longitudinal measurements are then simulated as $y_i(t) \sim \mathcal{N}(\eta_i(t), 0.7)$. 

The survival process is defined via the subject-specific hazard function $\lambda_i(t)$ defined in \eqref{eq: phm}, with the corresponding survival function $S_i(T) = \exp\left( -\int_0^T \lambda_i(u) \, du \right)$. To simulate event times, we follow the approach of \cite{austin2012generating}, drawing $p_i \sim \text{Uniform}(0,1)$ and solving for the latent event time  $\tilde{T}_i$ such that
$$p_i = \exp\left( -\int_0^{\tilde{T}_i} \lambda_i(u) \, du \right).$$ 
If the cumulative hazard function is linear in time, a closed-form solution for $\tilde{T}_i$ is available; otherwise, we apply the bisection method to numerically solve for $\tilde{T}_i$ within the interval $[0, 100]$. Censoring times $C_i$ are drawn from a uniform distribution over the latter half of the study period to reflect staggered entry and administrative censoring. The observed survival time is defined as $T_i = \min(\tilde{T}_i, C_i)$. For each subject, we identify $t_L$ as the largest time point in $t_{\text{grid}}$ less than $T_i$, and $t_U$ as the smallest time point greater than $T_i$. Events or censoring are assumed to occur within the interval $(t_L, t_U]$, and survival information is retained only up to $t_U$.

This entire simulation procedure is repeated independently for each subject. All unspecified parameters will be defined in the next section, corresponding to different settings and scenarios. Unless otherwise stated, a significance level of $0.05$ is used.

\subsection{Simulation Results}
Table \ref{tb:vssv} demonstrate that the sample size formula provides reliable estimates when the primary goal is to assess the overall SNP effect. Importantly, the power is largely unaffected by different random effect parameters $\delta^2_{0}$, $\delta^2_{1}$, and $\delta_{0,1}$ in the longitudinal process. The formula consistently approximates the required sample size across different values of $\alpha$, $\beta_g$, and $\gamma_g$, indicating its robustness in handling a variety of parameter settings.

\begin{table}[h!] 
\centering
\caption{Validation of sample size formula for testing the overall SNP effect for varying covariance structure, genetic effects, and association parameter}
\begin{tabularx}{\textwidth}{c *{10}{Y}}
\toprule
& & & & & & & & \multicolumn{2}{c}{Power} \\ \cmidrule(l){9-10}
$\alpha$ & $\beta_g$ & $\gamma_g$ & $\gamma_g + \alpha\beta_g$ & $\delta^2_{0}$ & $\delta^2_{1}$  & $\delta_{0, 1}$ & $\bar{D}$ & \scriptsize{Empirical} & \scriptsize{Calculated}\\
\midrule
0.25 & 0.3 & 0.1 & 0.175 & 2 & 1 & -0.1 & 610.21 & 0.787 & 0.800\\
0.25 & 0.3 & 0.1 & 0.175 & 2 & 0.5 & -0.6 & 616.49 & 0.798 & 0.804\\
0.25 & 0.3 & 0.1 & 0.175 & 2 & 0.5 & -0.1 & 610.21 & 0.801 & 0.800\\
0.25 & 0.3 & 0.1 & 0.175 & 2 & 0.1 & -0.3 & 608.66 & 0.802 & 0.799\\
0.25 & 0.3 & 0.1 & 0.175 & 2 & 0.1 & -0.1 & 607.12 & 0.809 & 0.798\\
\midrule
0.25 & 0.1 & 0.05 & 0.075 & 2 & 0.1 & -0.1 & 586.98 & 0.226 & 0.217\\
0.25 & 0.1 & 0.1 & 0.125 & 2 & 0.1 & -0.1 & 597.41 & 0.509 & 0.508\\
0.25 & 0.5 & 0.1 & 0.225 & 2  & 0.1 & -0.1 & 617.86 & 0.956 & 0.952\\
0.15 & 0.3 & 0.1 & 0.145 & 2 & 0.1 & -0.1 & 321.85 & 0.415 & 0.392\\ 
0.15 & 0.3 & 0.2 & 0.245 & 2 & 0.1 & -0.1 & 338.69 & 0.845 & 0.832\\
\bottomrule
\multicolumn{10}{l}{\footnotesize The longitudinal part is measured quarterly, and survival part is measured half yearly. The event time is simulated from} \\
\multicolumn{10}{l}{\footnotesize  Weibull distribution with $\lambda=0.01$, $v=1.1$; And $\beta_0 = 8.5$, $\beta_1 = 0.1$, $\beta_2 = 0$.}
\end{tabularx}
\label{tb:vssv}
\end{table}

Additionally, Table \ref{tb:vssn} shows that the sample size formula performs well across a wide range of significance levels. The formula remains relatively insensitive to the frequency of measurements (S1 – S4) as long as there are a sufficient number of measurements per subject. However, the performance of the formula declines when the number of measurements is very small. Specifically, comparing S4 and S5 highlights that power underestimation arises in cases of interval censoring, which affects the accuracy of the sample size estimates when fewer measurements are available. Overall, these simulation studies confirm that the sample size formula is effective in most scenarios but may require adjustments when dealing with limited data.

\begin{table}[h!]
\centering
\caption{Validation of sample size formula for testing the overall SNP effect for varying numbers of visits and significance levels}
\begin{tabularx}{\textwidth}{c *{7}{Y}}
\toprule
& \multicolumn{6}{c}{Power} \\ \cmidrule(l){2-7}
& Calculated & \multicolumn{5}{c}{Empirical} \\ \cmidrule(l){3-7}
Significance Level & & S1 & S2 &  S3  & S4 & S5\\
\midrule
$5\times10^{-2}$ & 0.978  & 0.961 & 0.968 & 0.958 & 0.946 & 0.908 \\
$10^{-2}$ & 0.919 & 0.891 & 0.903 & 0.880 & 0.863& 0.798 \\
$10^{-3}$  & 0.753 & 0.723 & 0.709 & 0.674 & 0.670 & 0.546\\
$10^{-4}$ & 0.533 & 0.551 & 0.508  & 0.495 & 0.442 & 0.312 \\
$10^{-5}$ & 0.329 & 0.359 & 0.349 & 0.300 & 0.270 & 0.161 \\
$10^{-6}$ & 0.179 & 0.211 & 0.202 & 0.167 & 0.141 & 0.072\\
$10^{-7}$ & 0.088 & 0.118 & 0.114 & 0.107 & 0.070 & 0.031\\
$10^{-8}$ & 0.039 & 0.067 & 0.061 & 0.054 & 0.026 & 0.008\\
\bottomrule
\multicolumn{7}{l}{\footnotesize  S1: Quarterly longitudinal visit \& half year time to event visit, S2: Half year longitudinal visit \& yearly time to event visit;} \\
\multicolumn{7}{l}{\footnotesize  S3: Yearly longitudinal visit \& six time to event visit, S4: Five longitudinal visit \& five time to event visit,  S5: Five } \\
\multicolumn{7}{l}{\footnotesize
longitudinal visit \& three time to event visit. The event time is simulated from Weibull distribution with $\lambda=0.001$, $v=1.1$;}\\
\multicolumn{7}{l}{\footnotesize  And $\beta_0 = 8.5$, $\beta_1 = 0.1$, $\beta_2=0$, $\alpha=0.5$, $\beta_g=0.3$, $\gamma_g=0.1$, $\delta^2_{0}=2$, $\delta^2_{1}=0.1$, $\delta_{0, 1}= -0.1$. The mean number of events} \\
\multicolumn{7}{l}{\footnotesize 
 observed in $1000$ simulation replicates, $\bar{D}$, is $601.64$ based on simulation.} 
\end{tabularx}
\label{tb:vssn}
\end{table}

The simulation studies presented in Table \ref{tb:vss} compare the power when the longitudinal model is misspecified. When the true quadratic model is correctly specified, the empirical power closely aligns with the calculated power, demonstrating that the model captures the underlying data structure accurately. However, when the longitudinal model is misspecified as linear, the empirical power deviates from the calculated power. This discrepancy arises because the true longitudinal trajectory exhibits additional curvature (i.e., an increase in $\beta_2$), which the linear model fails to capture. As a result, the linear model underestimates the complexity of the data, leading to a reduced empirical power compared to the calculated power. This highlights the importance of correctly specifying the longitudinal model to ensure accurate power estimation and reliable conclusions in hypothesis testing.

\begin{table}[h!] 
\centering
\caption{Comparison of empirical power to detect the overall SNP effect under misspecified longitudinal models, with $\beta_2$ representing the magnitude of nonlinear curvature in the longitudinal process}
\begin{tabularx}{\textwidth}{c *{9}{Y}}
\toprule
& & & & & &  \multicolumn{2}{c}{Power} \\ \cmidrule(l){7-9}
$\alpha$ & $\beta_g$ & $\gamma_g$ & $\gamma_g+\alpha\beta_g$ & $\bar{D}$ & $\beta_2$ & \scriptsize{Empirical} & \scriptsize{Empirical*} & \scriptsize{Calculated}\\
\midrule
0.35 & 0.2 & 0.05 & 0.12 & 426.91 & 0.1 & 0.355 & 0.351 & 0.362\\
0.35 & 0.2 & 0.05 & 0.12 & 562.47 & 0.15 & 0.427 & 0.411 & 0.454\\
0.35 & 0.2 & 0.05 & 0.12 & 689.97 & 0.2 & 0.515 & 0.461 & 0.533\\
\bottomrule
\multicolumn{9}{l}{\footnotesize The longitudinal part is measured quarterly, and survival part is measured half yearly. The event time is simulated from} \\
\multicolumn{9}{l}{\footnotesize   Weibull distribution with $\lambda=0.003$, $v=1.15$; $\beta_0 = 7.5$, $\beta_1 = -0.5$, $\delta^2_{0}=1$, $\delta^2_{1}=0.3$, $\delta^2_{2}=0.01$. Empirical is based on}\\
\multicolumn{9}{l}{\footnotesize fitting quadratic (true) longitudinal model, Empirical* is based on fitting linear (misspecified) longitudinal model.}
\end{tabularx}
\label{tb:vss}
\end{table}

\subsection{Biased estimates of the SNP effect when ignoring the longitudinal trajectory}
When a SNP influences the longitudinal process (i.e., $\beta_g \neq 0$ in \eqref{eq: eta}), and the longitudinal process in turn affects the time-to-event outcome (i.e., $\alpha \neq 0$ in \eqref{eq: phm}), the total effect of the SNP on the survival outcome is given by $\gamma_g + \alpha\beta_g$. It is well known that omitting the longitudinal process from the survival model in \eqref{eq: phm} can lead to biased estimation of $\gamma_g$. Even when the SNP is not associated with the longitudinal process (i.e., $\beta_g = 0$), failure to account for the longitudinal trajectory can still induce attenuation bias in estimating $\gamma_g$ due to unobserved heterogeneity \citep{horowitz1999semiparametric, abbring2007unobserved}.

Table \ref{tb:ae} presents the bias in estimating $\gamma_g$ under the null scenario where $\beta_g = 0$, comparing various modeling strategies: the Cox model that excludes the longitudinal process, two-stage methods (using either known or estimated trajectories), and the joint likelihood approach. The degree of bias is evaluated as a function of the association strength $\alpha$ between the longitudinal and survival processes. To ensure comparable event rates across scenarios, we set $\lambda = 0.01$ for $\alpha = 0.25$, and $\lambda = 0.001$ for $\alpha = 0.4$ and $0.5$. As $\alpha$ increases, the Cox model that omits the longitudinal process exhibits increasingly severe bias due to the unaccounted association. In contrast, both two-stage methods and the joint likelihood approach yield acceptable results. Interestingly, the two-stage methods slightly outperform the joint approach, aligning with prior findings that two-stage estimators can offer greater robustness under certain conditions \citep{Chen2011SampleData}.

\begin{table}[h!]
\centering
\caption{Effect of the value of $\alpha$ on the estimation of direct SNP effect $\gamma_g$ on survival based on different analysis methods: Separate, two stage and joint likelihood}
\begin{tabularx}{\textwidth}{c *{4}{Y}}
\toprule
& \multicolumn{3}{c}{Analysis model}\\
\cmidrule(l){2-5}
& \small $\lambda_i(t) = \lambda_0(t) \exp (\gamma_g \text{SNP}_i)$ & \multicolumn{3}{c}{\small $\lambda_i(t) = \lambda_0(t) \exp \lbrace \gamma_g \text{SNP}_i + \alpha((\beta_0 + b_{i0})+ (\beta_1+b_{i1}) t) \rbrace$} \\ \cmidrule(l){3-5}
$\alpha$ &  $\hat{\gamma}_g$ based on cox partial likelihood & $\hat{\gamma}_g$ based on known trajectory &$\hat{\gamma}_g$ based on Two Stage &$\hat{\gamma}_g$ based on joint likelihood\\
\midrule
0.25 & 0.093(0.002) & 0.101(0.002) & 0.101(0.002) & 0.096(0.002)\\
0.4 & 0.090(0.003) & 0.102(0.003) & 0.101(0.003) & 0.096(0.003)\\
0.5 & 0.077(0.002) & 0.100(0.002) & 0.100(0.002) & 0.095(0.002)\\
\bottomrule
\multicolumn{5}{l}{\footnotesize The longitudinal part is measured quarterly, and survival part is measured half yearly. The event time is simulated from} \\
\multicolumn{5}{l}{\footnotesize Weibull distribution with $\lambda$, $v=1.1$; And $\beta_0 = 8.5$, $\beta_1 = 0.1$, $\delta^2_{0}$ = 2, $\delta^2_{1}$ =0.1, $\delta_{0, 1}= -0.1$. True $\gamma_g = 0.1$. And $\hat{\gamma}_g$ is}\\
\multicolumn{5}{l}{\footnotesize summarized as mean (SD) based on 1000 simulations of 1000 subjects. $\hat{\gamma}_g$.}
\end{tabularx}
\label{tb:ae}
\end{table}

\subsection{Relationship between the calculated power and the sample size, maximum follow-up time, and other key parameters}
\begin{figure}[h]
\centering
\includegraphics[width=\linewidth]{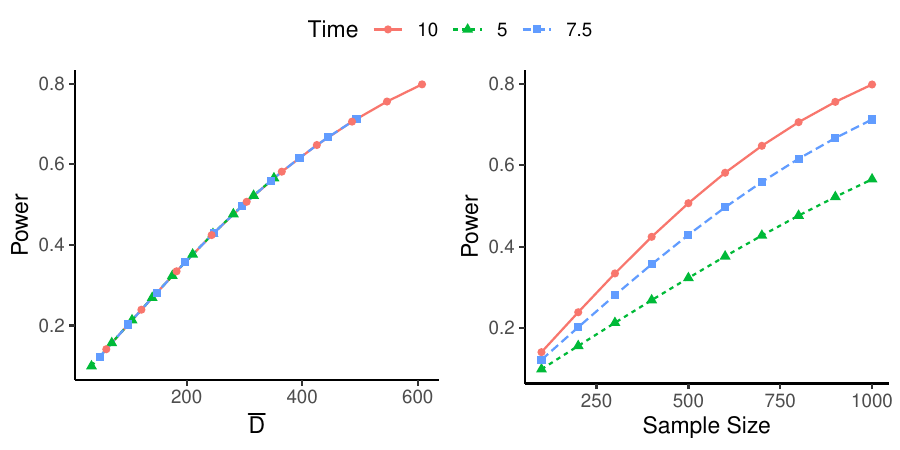}
\caption{Power curve at different follow-up times}
\label{fig: p.f}
\end{figure}

\begin{figure}[h!] 
    \centering
  \includegraphics[width=\linewidth]{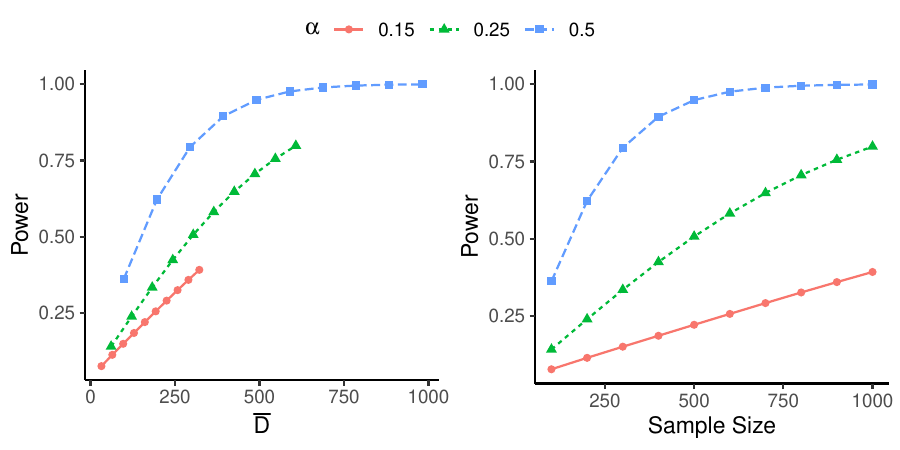}
  \caption{Power curve when only $\alpha$ varies}
  \label{fig: p.a}
\end{figure}

\begin{figure}[h!]
  \centering
  \includegraphics[width=\linewidth]{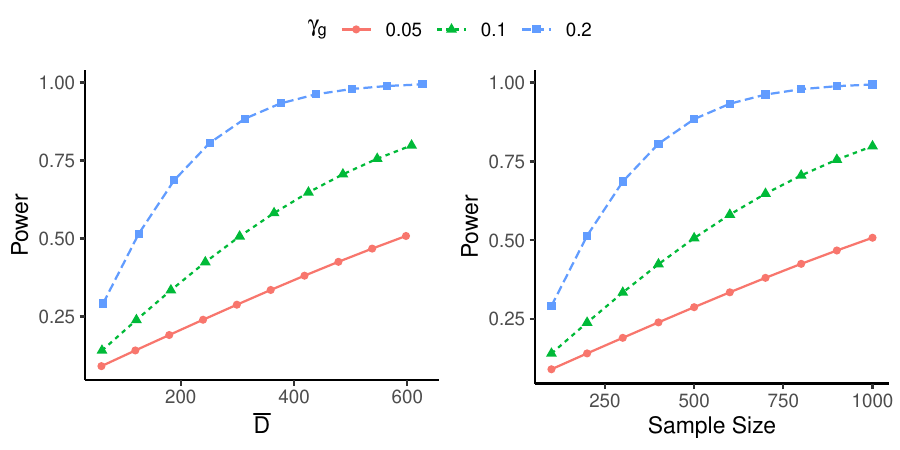}
  \caption{Power curve when only $\gamma_g$ varies}
  \label{fig: p.g}
\end{figure}

We further investigate how the calculated power varies with sample size, event size, maximum follow-up time, and other key parameters using the proposed sample size formula for the overall SNP effect. Unless otherwise specified, the maximum follow-up time is set to 10 years, with longitudinal measurements collected quarterly and survival outcomes assessed semi-annually; event times are simulated from a Weibull distribution with shape parameter \( v = 1.1 \) and scale parameter \( \lambda = 0.01 \). The following model parameters are used: \( \beta_0 = 8.5 \), \( \beta_1 = 0.1 \), \( p = 0.3 \), \( \delta_0^2 = 2 \), \( \delta_1^2 = 0.1 \), \( \delta_{0,1} = -0.1 \), \( \gamma_g = 0.1 \), \( \alpha = 0.25 \), and \( \beta_g = 0.3 \). The sample sizes and corresponding estimated numbers of events, denoted by $\bar{D}$, are summarized in Tables \ref{tab: p.f}, \ref{tab: p.a}, and \ref{tab: p.g}, respectively. Figures \ref{fig: p.f}, \ref{fig: p.a}, and \ref{fig: p.g} display the calculated power as a function of $\bar{D}$ and sample size, across different follow-up times and varying values of $\alpha$ and $\gamma_g$.

Figure \ref{fig: p.f} presents the power curves for the overall SNP effect at three follow-up times: 5, 7.5, and 10 years. The results reveal a clear pattern: for a fixed sample size, longer follow-up times yield higher power. This is because extended follow-up allows for the accumulation of more survival data through additional observed events. As more information becomes available over time, the statistical power to detect the effect increases. When plotting power against $\bar{D}$, the curves collapse into a single trajectory, consistent with the relationship described in \eqref{eq: p}.

\begin{table}[h!]
    \centering
    \caption{Corresponding $\bar{D}$ values for Figure \ref{fig: p.f}}
    \label{tab: p.f}
    \begin{tabular}{c c c c c c c c c c c}
    \toprule
    Followup & \multicolumn{10}{c}{Sample Size}\\
    \midrule
    & 100 & 200 & 300 & 400 & 500 & 600 & 700 & 800 & 900 & 1000\\
    \midrule
    5 & 35.26 & 70.34 & 105.46 & 140.33 & 175.24 & 210.38 & 245.54 & 280.86 & 315.88 & 351.15\\
    7.5 & 49.36 & 98.67 & 148.12 & 197.35 & 246.71 & 296.15 & 345.663 & 395.25 & 444.49 & 493.90\\
    10 &  60.86 & 121.66 & 182.52 & 243.30 & 303.91 & 364.69 & 425.38 & 486.28 & 546.85 & 607.51\\
    \bottomrule
    \multicolumn{11}{l}{\footnotesize D is the mean number of events observed in simulation replicates.}
    \end{tabular}
\end{table}

Figures \ref{fig: p.a} and \ref{fig: p.g} further illustrate how both $\alpha$ and $\gamma_g$ affect $\bar{D}$ and, consequently, power. Figure \ref{fig: p.a} shows power curves for three values of $\alpha$: 0.15, 0.25, and 0.5. As expected, power decreases with smaller $\alpha$ at the same sample size, reflecting the fact that weaker effects require larger samples to achieve the same power. When power is plotted against $\bar{D}$, the curves no longer align due to variation in both $\bar{D}$ and $\alpha$ in \eqref{eq: p}. The power still declines with decreasing $\alpha$ for a given $\bar{D}$. Similarly, Figure \ref{fig: p.g} displays power curves for three values of $\gamma_g$: 0.05, 0.1, and 0.2. As with $\alpha$, smaller values of $\gamma_g$ result in lower power at the same sample size or $\bar{D}$, again underscoring that smaller effect sizes necessitate larger samples. The patterns observed in Figures \ref{fig: p.a} and \ref{fig: p.g} are analogous, as both $\alpha$ and $\gamma_g$ contribute to the overall effect size, which in turn influences the sample size formula.

\begin{table}[h!]
    \centering
    \caption{Corresponding $\bar{D}$ values for Figure \ref{fig: p.a}}
    \label{tab: p.a}
    \begin{tabular}{c c c c c c c c c c c}
    \toprule
    $\alpha$ & \multicolumn{10}{c}{Sample Size}\\
    \midrule
    & 100 & 200 & 300 & 400 & 500 & 600 & 700 & 800 & 900 & 1000\\
    \midrule
    0.15 & 32.30 & 64.46 & 96.62 & 128.54 & 160.68 & 192.87 & 225.24 & 257.56 & 289.70 & 321.87\\
    0.25 & 60.86 & 121.66 & 182.52 & 243.30 & 303.91 & 364.69 & 425.38 & 486.28 & 546.85 & 607.51\\
    0.5 & 98.25 & 196.49 & 294.71 & 392.95 & 491.22 & 589.51 & 687.72 & 785.94 & 884.15 & 982.35\\
    \bottomrule
    \multicolumn{11}{l}{\footnotesize D is the mean number of events observed in simulation replicates.}
    \end{tabular}
\end{table}

\begin{table}[h!]
    \centering
    \caption{Corresponding $\bar{D}$ values for Figure \ref{fig: p.g}}
    \label{tab: p.g}
    \begin{tabular}{c c c c c c c c c c c}
    \toprule
    $\gamma_g$ & \multicolumn{10}{c}{Sample Size}\\
    \midrule
    & 100 & 200 & 300 & 400 & 500 & 600 & 700 & 800 & 900 & 1000\\
    \midrule
    0.05 & 59.89 & 119.76 & 179.57 & 239.35 & 299.01 & 358.77 & 418.46 & 478.38 & 537.97 & 597.63\\
    0.1 & 60.86 & 121.66 & 182.52 & 243.30 & 303.91 & 364.69 & 425.38 & 486.28 & 546.85 & 607.51\\
    0.2 & 62.79 & 125.47 & 188.23 & 250.89 & 313.47 & 376.18 & 438.83 & 501.68 & 564.20 & 626.85\\
    \bottomrule
    \multicolumn{11}{l}{\footnotesize D is the mean number of events observed in simulation replicates.}
    \end{tabular}
\end{table}

\section{Retrospective Power Analyses for the Diabetes Control and Complications Trial}\label{sec: a}
To demonstrate the application of the proposed sample size formula, we conducted a retrospective power analysis using data from the Diabetes Control and Complications Trial (DCCT; \citealp{TheDCCTResearchGroup1986,TheDCCTResearchGroup1990}). The DCCT was a prospective, randomized controlled clinical trial designed to compare the effects of intensive versus conventional treatment on the onset and progression of long-term microvascular and neurological complications in individuals with type 1 diabetes. Under conventional treatment, patients received one or two daily injections of a combination of intermediate- and short-acting insulin. They performed daily self-monitoring of urine or blood glucose levels, received education on diet and exercise, but typically did not adjust insulin dosages on a daily basis. In contrast, patients in the intensive treatment arm received insulin three or more times daily via injection or external pump. Insulin doses were adjusted based on self-monitoring of blood glucose levels performed at least four times per day, dietary intake, and anticipated physical activity. The treatment objective in the intensive arm was to maintain HbA1c levels within the normal range, defined as less than 6.05 percent \citep{dcct1993}.

Over a follow-up period ranging from three to nine years, with a mean duration of 6.5 years, the conventional arm maintained an average hemoglobin A1C (HbA1c) level of approximately 9.0 percent, consistent with their baseline values. The intensive arm achieved and sustained a mean HbA1c level of approximately 7.0 percent. HbA1c, a measure of long-term glycemic exposure, was collected quarterly in the conventional arm and monthly in the intensive arm. Subjects in the conventional arm had an average of 17.26 quarterly measurements, with a minimum of 3, a median of 17.5, and a maximum of 37. Subjects in the intensive arm had an average of 18.35 quarterly measurements, with a minimum of 3, a median of 19, and a maximum of 38.

Clinical complications were assessed through periodic evaluations, with retinopathy status determined by grading retinal photographs taken at semiannual visits over a maximum duration of ten years. Retinopathy was selected as the primary outcome due to its well-characterized natural history and reliable assessment \citep{TheDCCTResearchGroup1990}. The primary endpoint for this analysis is the time from study entry to the onset of mild retinopathy. Subjects with retinopathy diagnosed at baseline were excluded for our analysis. Censoring was administrative and determined by a predefined calendar date.

With the development of genome-wide association study (GWAS) methodologies, DCCT participants underwent genotyping using the Illumina 1M beadchip assay. After quality control procedures, a total of 841,342 SNPs with minor allele frequency greater than one percent were retained for analysis. \citet{Paterson2010} identified a major locus near \textit{SORCS1} on chromosome 10q25.1 that was significantly associated with HbA1c levels in the conventional arm, with replication in the intensive arm. Other loci nearing genome-wide significance included regions at 14q32.13 (\textit{GSC}) and 9p22 (\textit{BNC2}) in the combined treatment arms, and 15q21.3 (\textit{WDR72}) in the intensive arm. Some of these loci were also associated with diabetic complications, notably \textit{BNC2}, which was linked to both renal and retinal outcomes. The direction of association at this locus was consistent with its effect on HbA1c: genotypes associated with higher HbA1c levels were also associated with increased complication risk.

Due to substantive differences in treatment protocols and outcomes, and in accordance with most prior analyses of the DCCT, we conducted separate power analyses for each treatment arm. The conventional treatment group included 526 subjects, among whom 315 experienced the event of interest. The intensive treatment group included 527 subjects, with 232 observed events. 

\begin{figure}[h!]
  \centering
  \includegraphics[width=\linewidth]{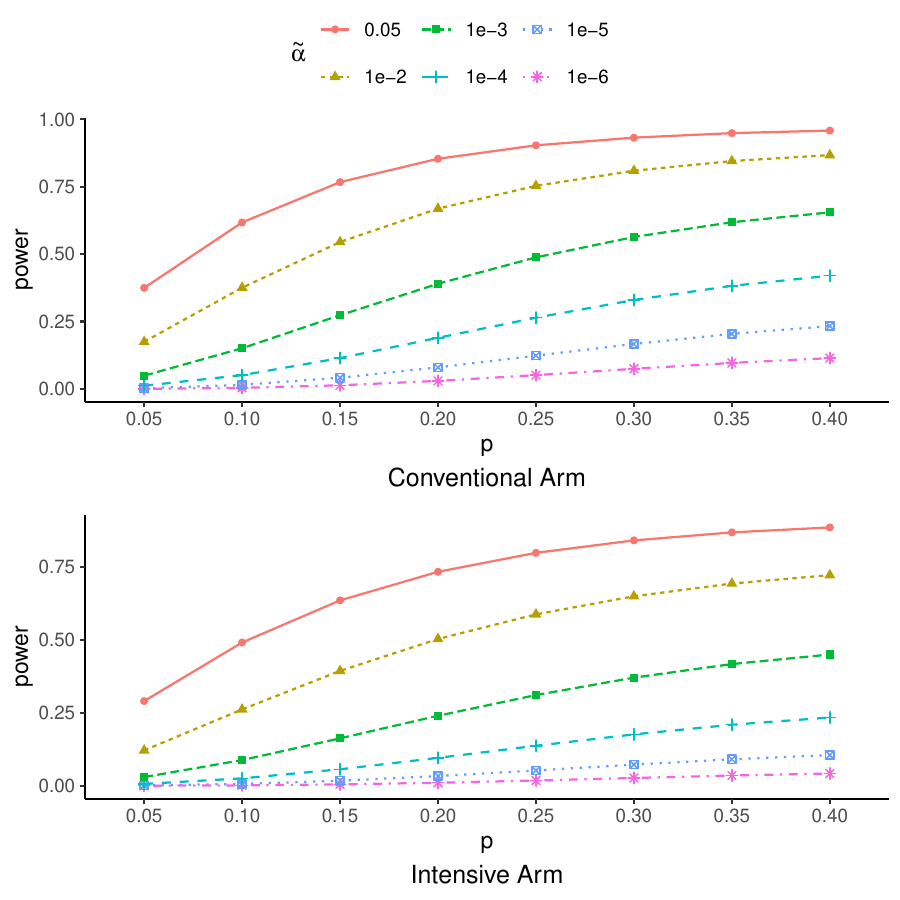}
  \caption{Power curve when $p$ and $\tilde{\alpha}$ vary}
  \label{fig: rpa}
\end{figure}

Figure \ref{fig: rpa} presents power curves for both the conventional and intensive treatment arms, assuming an overall SNP effect size of $\theta_g = 0.30$. The plots demonstrate how statistical power depends jointly on the minor allele frequency $p$ and the significance level $\tilde{\alpha}$. In the conventional arm, 80\% power is achieved when $p > 0.20$ under the standard significance level of $\tilde{\alpha} = 0.05$. In contrast, for the intensive arm, where the number of events is lower, the same power threshold is only reached when $p \approx 0.40$ at $\tilde{\alpha} = 0.05$. These results indicate that the DCCT sample size is insufficient to attain adequate power under conventional GWAS significance thresholds in either treatment arm. Notably, the intensive arm exhibits greater sensitivity to both allele frequency and the choice of $\tilde{\alpha}$, reflecting the compounding effects of reduced sample size and event rate on statistical power. This underscores the importance of considering study design factors, particularly sample size and event accrual, when assessing the feasibility of detecting modest genetic effects in stratified analyses.

\section{Discussion}\label{sec: d}

In this paper, we propose a sample size formula specifically designed for estimating the overall SNP effect within a joint modeling framework. Our empirical results show that the formula provides accurate approximations across a wide range of scenarios, including varying magnitudes of the SNP effect, different visit schedules, random effect structures, and significance levels. However, when the number of visit schedules is too small, the accuracy of the sample size estimates deteriorates. Additionally, we show that inverting the sample size formula to estimate the power for detecting overall genetic effects performs well in simulation studies. To demonstrate practical applicability, we conducted a retrospective power analysis using data from the DCCT study.

Notably, although we assume that the random effects and error terms in the longitudinal submodel follow a normal distribution, the validity of the sample size formula itself does not rely on this assumption. As demonstrated by \cite{Chen2011SampleData}, the formula remains applicable in more general joint modeling settings, thereby enhancing its practical utility in a wide range of genetic association studies involving longitudinal and time-to-event data. Two key assumptions underlie the derivation and application of our sample size formula. First, it assumes that the longitudinal process is subject to non-informative censoring. This is a critical condition, as the two-stage estimation procedure used in our derivation can produce biased results under informative censoring. Therefore, caution is warranted when applying the formula in settings where informative dropout may be present. Second, when the longitudinal trait functions as a biomarker, we assume that the indirect genetic effect on the survival outcome $\alpha \beta_g$ and the direct effect on the survival trait $\gamma_g$ act in the same direction. Violating this assumption may impact both the interpretation of genetic effects and the accuracy of power estimation.

In practice, the longitudinal submodel is often nonlinear and unknown, and the accuracy of the sample size formula may be influenced by how well the longitudinal process is specified. For instance, as demonstrated by \cite{bian2020hypothesis}, visualizations of the DCCT data reveal distinct trajectories between subjects in the control and intensive treatment arms; and misspecification of the longitudinal process can lead to incorrect parameter estimation. Such differences emphasize the importance of correctly modeling the longitudinal trait, as misspecification can lead to biased parameter estimates, affecting both the power analysis and genetic effect estimation. This highlights a key challenge in applying the proposed sample size formula in real-world studies: the model's sensitivity to inaccuracies in the specification of longitudinal trajectories. Further exploration is needed to assess how model misspecifications, such as incorrect assumptions about the trajectories or error structures, may influence the validity of power calculations and subsequent genetic association findings. Moreover, it would be valuable to investigate strategies for improving the robustness of the sample size formula when faced with model misspecification, including alternative approaches that can better accommodate variations in longitudinal trajectories across different treatment groups or patient populations.

\section*{Acknowledgements}
This research is partially supported by a Canadian Institutes of Health Research (CIHR) project grant.

\bibliographystyle{apalike}
\bibliography{references}

\end{document}